\documentclass{fic-l}

\usepackage{amssymb}
\input diagrams         

\newtheorem{theorem}{Theorem}[section]
\newtheorem{lemma}[theorem]{Lemma}

\theoremstyle{definition}
\newtheorem{definition}[theorem]{Definition}

\theoremstyle{remark}
\newtheorem{remark}[theorem]{Remark}

\newtheorem{proposition}[theorem]{Proposition}
\newtheorem{corollary}[theorem]{Corollary}

\newtheorem{conjecture}[theorem]{Conjecture}

\numberwithin{equation}{section}

\def\1#1{{\bf #1}}
\def\2#1{{\mathcal #1}}
\def\3#1{{\sl #1}}
\def\4#1{{\tt #1}}
\def\5#1{{\sf #1}}
\def\6#1{{\mathfrak #1}}
\def\7#1{{\mathbb #1}}

\newcommand{\prf}{{\it Proof. }}
\renewcommand{\qed}{\ \hfill $\blacksquare$}

\newcommand{\bdefin}{\begin{definition}}
\newcommand{\blemma}{\begin{lemma}}
\newcommand{\bprop}{\begin{proposition}}
\newcommand{\btheor}{\begin{theorem}}
\newcommand{\bcoro}{\begin{corollary}}
\newcommand{\bconj}{\begin{conjecture}}
\newcommand{\edefin}{\end{definition}}
\newcommand{\elemma}{\end{lemma}}
\newcommand{\eprop}{\end{proposition}}
\newcommand{\etheor}{\end{theorem}}
\newcommand{\ecoro}{\end{corollary}}
\newcommand{\econj}{\end{conjecture}}
\newcommand{\brem}{\begin{remark}}
\newcommand{\erem}{\end{remark}}

\newcommand{\ba}{\begin{array}}
\newcommand{\ea}{\end{array}}
\newcommand{\bea}{\begin{eqnarray}}
\newcommand{\eea}{\end{eqnarray}}
\newcommand{\bean}{\begin{eqnarray*}}
\newcommand{\eean}{\end{eqnarray*}}
\newcommand{\nn}{\nonumber}

\newcommand{\impl}{\Rightarrow}
\newcommand{\rarr}{\rightarrow}
\newcommand{\restr}{\upharpoonright}
\newcommand{\ol}{\overline}
\newcommand{\ul}{\underline}
\newcommand{\obj}{\mbox{Obj}}
\newcommand{\Hom}{\mbox{Hom}}

\newcommand{\id}{\mbox{id}}
\renewcommand{\mod}{\mbox{mod}}

\begin{document}

\title{Conformal Field Theory and \\ Doplicher-Roberts Reconstruction}

\author{Michael M\"uger}

\address{School of Mathematical Sciences, Tel Aviv University, Ramat Aviv, Israel}


\email{mueger@x4u2.desy.de}

\thanks{Financially supported by the European Union through the TMR Networks
`Noncommutative Geometry' and `Orbits, Crystals and Representation Theory'}

\subjclass[2000]{Primary: 81T40. Secondary: 81T05, 46L60, 18D10}

\date{August 20, 2000}

\dedicatory{This contribution is dedicated to Sergio Doplicher and John E.\ Roberts\\
on the occasion of their 60th birthdays.}

\begin{abstract}
After a brief review of recent rigorous results concerning the representation theory of
rational chiral conformal field theories (RCQFTs) we focus on pairs $(\2A, \2F)$ of 
conformal field theories, where $\2F$ has a finite group $G$ of global symmetries and
$\2A$ is the fixpoint theory. The comparison of the representation categories of $\2A$ and
$\2F$ is strongly intertwined with various issues related to braided tensor categories. We
explain that, given the representation category of $\2A$, the representation category of
$\2F$ can be computed (up to equivalence) by a purely categorical construction. The latter
is of considerable independent interest since it amounts to a Galois theory for braided
tensor categories. We emphasize the characterization of modular categories as braided
tensor categories with trivial center and we state a double commutant theorem for
subcategories of modular categories. The latter implies that a modular category $\2M$
which has a replete full modular subcategory $\2M_1$ factorizes as
$\2M\simeq\2M_1\otimes_\7C\2M_2$ where $\2M_2=\2M\cap\2M_1'$ is another modular
subcategory. On the other hand, the representation category of $\2A$ is not determined
completely by that of $\2F$ and we identify the needed additional data in terms of soliton
representations. We comment on `holomorphic orbi\-fold' theories, i.e.\ the case where $\2F$
has trivial representation theory, and close with some open problems.

We point out that our approach permits the proof of many conjectures and heuristic results
on `simple current extensions' and `holomorphic orbi\-fold models' in the physics literature
on conformal field theory.
\end{abstract}

\maketitle

\section{Introduction}
As is well known and will be reviewed briefly in the next section, quantum field theories
in Minkowski space of not too low dimension give rise to representation categories which
are symmetric $C^*$-tensor categories with duals and simple unit. (The minimum number of
space dimensions for this to be true depends on the class of representations under
consideration.) As Doplicher and Roberts have shown, such categories are representation
categories of compact groups \cite{dr6} and every QFT is the fixpoint theory under a compact
group action \cite{dr2} of a theory admitting only the vacuum representation \cite{cdr}.
Thus the theory of (localized) representations of QFTs in higher dimensional spacetimes is
essentially closed.

Though this is still far from being the case for low dimensional theories there has been
considerable recent progress, of which we will review two aspects. The first of these
concerns the general representation theory of rational chiral conformal theories, which
have been shown \cite{klm} to give rise to unitary modular categories in perfect
concordance with the physical expectations. See also \cite{mue13} for a more
self-contained and (somewhat) more accessible review. In this contribution we restrict
ourselves to stating the main results insofar as they serve to motivate the subsequent
considerations which form the core of this paper. 

We will then study pairs $(\2F, \2A)$ of quantum field theories in  low dimension,
mostly rational conformal, where $\2A$ is the fixpoint theory of $\2F$ w.r.t.\ the action
of a finite group $G$ of global symmetries. This scenario may seem quite special, as in fact
it is, but it is justified by several arguments. First of all, as already alluded to, the
fixpoint situation is the generic one in high dimensions. Whereas this is definitely not
true in the case at hand, every attempt at classifying rational conformal field theories
(or at least modular categories) will most likely make use of constructions which
produce new conformal field theories from given ones. (Besides those we focus on there
are, of course, other such procedures like the `coset construction'.)
The converse of the passage to $G$-fixpoints is provided by the construction \cite{dr2}
of Doplicher and Roberts, which in the case of abelian groups has appeared in the CQFT
literature as `simple current extension'. The latter are of considerable relevance in the
classification of `modular invariants', i.e.\ the construction of two-dimensional CQFTs
out of chiral ones. It is therefore very satisfactory that we are able to provide rigorous
proofs for many results in this area.

Finally the analysis of quantum field theories related by finite groups leads to many
mathematical results which can be phrased in a purely categorical manner. As such they
have applications to other areas of mathematics like subfactor theory or low-dimensional
topology.

\section{`Many' Spacetime Dimensions: Symmetric Categories}
\subsection{Global Symmetry Groups in $\ge 2+1$ Spacetime Dimensions} \label{high_dim}
In this section we consider quantum field theories in Minkowski space with $d=s+1$
dimensions where the {\it number $s$ of space dimensions is at least two}. 
(See \cite{haag, K} for more details.) We denote by $\2K$ the set of double cones. 
Let $\2O\mapsto\2F(\2O), \2O\in\2K$ be a net (inclusion preserving assignment) of von Neumann
algebras on a Hilbert space $\2H$ satisfying irreducibility, locality 
($\2O_1\subset\2O_2' \impl [\2F(\2O_1),\2F(\2O_2)]=\{0\}$) and covariance w.r.t.\
a positive energy representation of the Poincar\'{e} group with invariant vacuum vector
$\Omega$. We sharpen the locality requirement by imposing Haag duality
\[ \2F(\2O)'=\2F(\2O')^{\ol{\ }} \quad\mbox{where} \quad \2F(\2O') =
  \ol{\cup_{\tilde{\2O}\in\2K, \tilde{\2O}\subset\2O'}  \2F(\tilde{\2O})}^{\|\cdot\|}.
\]
We assume that there is a compact group $G$ with a strongly continuous faithful unitary
representation $U$ on $\2H$ commuting with the representation of the Poincar\'{e} group,
leaving $\Omega$ invariant and implementing global symmetries of $\2F$:
\[ \alpha_g(\2F(\2O))=\2F(\2O) \ \ \forall g\in G, \2O\in\2K
   \quad\mbox{where}\ \alpha_g=Ad\,U(g). 
\]
Consider the subnet $\2A(\2O)=\2F(\2O)^G$ together with its vacuum representation $\pi_0$
on the subspace $\2H_0$ of $G$-invariant vectors. $\pi_0$ can be shown to satisfy Haag
duality \cite{dhr1}. The Hilbert space $\2H$ decomposes as
\[ \2H=\bigoplus_{\xi\in\hat{G}} \2H_\xi \otimes \7C^{d(\xi)}, \]
where $\hat{G}$ is the set of isomorphism classes of irreducible representation of $G$
and the group $G$ and the $C^*$-algebra 
$\2A=\ol{\cup_{\2O\in\2K}\2A(\2O)}^{\|\cdot\|}$ (the `quasi-local algebra')
act on $\2H$ as follows:
\[ U(g)=\bigoplus_{\xi\in\hat{G}} \11 \otimes U_\xi(g), \quad
   \pi(A)= \bigoplus_{\xi\in\hat{G}} \pi_\xi(A) \otimes \11, \quad g\in G, A\in\2A. \]
The representations $\pi_\xi$ of $\2A$ on $\2H_\xi$ are irreducible and satisfy \cite{dhr1}
\begin{equation} \pi_\xi \restr \2A(\2O') \cong \pi_0 \restr \2A(\2O')
  \quad\forall\2O\in\2K, 
\label{DHR}\end{equation}
where $\pi_0$ is the representation of $\2A$ on $\2H_0$.

These observations motivate the analysis of the positive energy representations satisfying
the `DHR criterion' (\ref{DHR}) for any irreducible local net $\2A$ of algebras satisfying
Haag duality and Poincar\'{e} covariance. 

\bdefin Let $\2O\mapsto\2A(\2O)\subset\2B(\2H_0)$ be a Haag dual net of algebras which is 
covariant w.r.t.\ a positive energy vacuum representation of $\2P$. Then
$\widetilde{DHR}(\2A)$ is the category of Poincar\'{e} covariant representations satisfying
(\ref{DHR}) ($\pi_0$ is the defining representation) together with their (bounded)
intertwining operators.
\edefin

For the purposes of the development of the theory another category is much more
convenient. 
\bdefin 
Let $\2A$ be as above. Then $DHR(\2A)$ denotes the category of localized
transportable morphisms, i.e.\ bounded unital $*$-algebra endomorphisms $\rho$ of the
quasi-local algebra $\2A$ such that $\rho\restr\2A(\2O')=\id$ for {\it some} 
$\2O\in\2K$ and such that for {\it every} $\tilde{\2O}\in\2K$ there is
$\rho_{\tilde{\2O}}$ localized in $\tilde{\2O}$ such that $\rho$ and $\rho_{\tilde{\2O}}$
are inner equivalent. The morphisms are the intertwiners in $\2A$.
\edefin

\btheor \label{thm1} \cite{dhr3} 
$DHR(\2A)$ is canonically isomorphic to a full subcategory of $\widetilde{DHR}(\2A)$ which
is equivalent to $\widetilde{DHR}(\2A)$. $DHR(\2A)$ is a strict symmetric $C^*$-tensor
category with simple tensor unit. There is an additive and multiplicative dimension
function on the objects with values in $\7N\cup{\infty}$. (If the 
local algebras $\2A(\2O), \2O\in\2K$ are assumed to be factors then
$d(\rho)^2=[\2A(\2O):\rho(\2A(\2O))]$ if $\rho$ is localized in $\2O$.) 
The full monoidal subcategory $DHR_f(\2A)$ of the $\rho$ with finite statistics
(i.e.\ $d(\rho)<\infty$) has conjugates in the sense of \cite{lro}. Viz. for every
$\rho\in DHR_f(\2A)$ there are $\ol{\rho}\in DHR_f(\2A)$ and
$r_\rho\in\Hom(\11,\ol{\rho}\rho)$, $\ol{r}_\rho\in\Hom(\11,\rho\ol{\rho})$ satisfying 
$r_\rho^*\circ r_\rho=\ol{r}_\rho^*\circ\ol{r}_\rho=d(\rho)\id_\11$ and
\[ \id_\rho\otimes r^*_\rho \,\circ\, \ol{r}_\rho\otimes\id_\rho =\id_\rho, \quad
  \id_{\ol{\rho}}\otimes \ol{r}^*_\rho \,\circ\, r_\rho\otimes\id_{\ol{\rho}}
   =\id_{\ol{\rho}}. \]
$DHR_f(\2A)$ is a ribbon category, i.e.\ has a twist, which on simple objects takes
the values $\pm 1$ (Bose-Fermi alternative). 
\etheor

Applying the general formalism to fixpoint nets as above one obtains:

\bprop \cite{dhr1} Let $\2A$ be a fixpoint net as above. Then $DHR_f(\2A)$ contains a
full monoidal subcategory $\2S$ which is equivalent (as a symmetric monoidal category) to the
category $G-\mod$ of finite dimensional continuous unitary representations of $G$. For a
simple object $\rho_\xi\in\2S$ the dimension $d(\rho_\xi)$ coincides with the dimension
$d(\xi)$ of the associated representation $U_\xi$ of $G$.
\eprop

Now the question arises under which circumstances one obtains all DHR representations of
$\2A$ in this way.

\bprop \cite{rob} \label{roberts}
Assume  $\2F$ has trivial representation category $DHR(\2F)$ (in
the sense of quasi-trivial one-cohomology). Then $\2A$ has no irreducible DHR
representations of infinite statistics and $DHR(\2A)\simeq G-\mod$.
\eprop

It is thus natural to conjecture that every net $\2A$ satisfying the above axioms is 
the fixpoint net under the action of a compact group $G$ of a net $\2F$ with trivial 
representation structure.

\subsection{The Reconstruction Theory of Doplicher and Roberts}\label{doro}
\btheor \cite{dr6} Let $\2S$ be a symmetric $C^*$-tensor category with conjugates and
simple unit such that every simple object has twist $+1$. Then there is a compact group
$G$, unique up to isomorphism, such that one has an equivalence $\2S\simeq G-\mod$ of
symmetric tensor $*$-categories with conjugates.
\label{DR-1}\etheor

\brem 1. If there are objects with twist $-1$ then there is a compact group $G$ together with
a central element $k$ of order two such that $\2S\simeq G-\mod$ as a tensor category and
the twist of a simple object equals the value of $k$ in the corresponding irreducible
representation of $G$.

2. Most categories in this paper will be closed w.r.t.\ direct sums and subobjects (i.e.\
all idempotents split). Yet, in order not to have to require this everywhere, all
equivalences of ((braided/symmetric) monoidal) categories in this paper will be understood
as equivalences of the respective categories after completion w.r.t.\ direct sums and
subobjects. See, e.g., \cite{gr} for these constructions and note that equivalence of the
completed categories is equivalent to Morita equivalence \cite{gr}. We believe that this
is the appropriate notion of equivalence for semisimple $k$-linear categories.
By the coherence theorem for braided tensor categories \cite{js1} we may and do assume
that all tensor categories are strict. (In fact, most of the categories under
consideration here are so by construction.) 
\erem

\btheor \cite{dr2} Let $\2O\mapsto\2A(\2O)\subset\2B(\2H_0)$ as above. Then there is a net
of algebras $\2O\mapsto\2F(\2O)\subset\2B(\2H)$ where $\2H\supset\2H_0$ such that 
\begin{itemize}
\item $\2F$ is a graded local net (which is local iff all objects in $DHR_f(\2A)$ have
twist $+1$), 
\item the group $G$ corresponding to the symmetric tensor category $DHR_f(\2A)$ is
unitarily and faithfully represented on $\2H$, implementing global symmetries of $\2F$,
\item $\2F(\2O)^G\restr\2H_0=\2A(\2O)\ \ \forall\2O\in\2K$,
\item the reducible representation of $\2A$ on $\2H$ contains every irreducible DHR sector
$\pi_\xi$ of $\2A$ (of finite dimension $d(\pi_\xi)$) with multiplicity $d(\pi_\xi)$,
\item the charged (non-$G$-invariant) fields intertwine the vacuum and the DHR sectors.
\end{itemize}
The net $\2F$, which we denote $\2F=\2A\rtimes DHR_f(\2A)$, is unique up to unitary
equivalence. (One may also consider the crossed product $\2A\rtimes\2S$ with a full
monoidal subcategory $\2S$ of $DHR_f(\2A)$.)
\label{DR-2}\etheor

It is natural to ask whether there is a converse to Prop.\ \ref{roberts} to the effect
that $\2F=\2A\rtimes DHR_f(\2A)$ has trivial representation theory. A first result was
proved independently in \cite{conti} and \cite{mue4}:

\bprop Assume that $\2A$ has finitely many unitary equivalence classes of irreducible DHR 
representations of finite statistics, all with twist $+1$. Then the local net
$\2F=\2A\rtimes DHR_f(\2A)$ has no non-trivial DHR representations of finite statistics.
\eprop 

This result has the obvious weakness of being restricted to theories with finite
representation theory. On the positive side, we do not need to make assumptions on
potential representations of $\2A$ with infinite statistics. For most purposes of the
present paper this result is sufficient, but we cite the following recent result.

\btheor \cite{cdr} Assume $\2A$ lives on a separable Hilbert space and all DHR
representations are direct sums of irreducible DHR representations with finite
statistics and twist $+1$. Then $\2F=\2A\rtimes DHR(\2A)$ has no non-trivial sectors of
finite or infinite statistics. 
\etheor

In $1+1$-dimensional Minkowski space or on $\7R$ (i.e.\ no time: `$1+0$ dimensions') the
DHR analysis must be modified \cite{frs1} since there one can only prove that $DHR(\2A)$
is braided. We will therefore give a brief discussion of some pertinent results on
braided tensor categories. (See \cite{js1} or \cite{ka} for the basic definitions.)

\section{Few Spacetime Dimensions and Modular Categories}
\subsection{Categorical Interlude 1: Braided Tensor Categories and Their Center}\label{Int-1}
Throughout we denote morphisms in a category by small Latin letters and objects by capital
Latin or, in the quantum field context, by small Greek letters. We often write $XY$
instead of $X\otimes Y$.

\bdefin A T$C^*$ is a $C^*$-tensor category \cite{lro} with simple unit and conjugates
(and therefore finite dimensional hom-spaces). A BT$C^*$ is a T$C^*$ with unitary braiding. A
ST$C^*$ is a symmetric BT$C^*$.
\edefin

A T$C^*$ (more generally, a semisimple spherical category) will be called finite
dimensional if the set $\Gamma$ of isomorphism classes of simple objects is finite.
Then its dimension is defined by 
\[ \dim \2C \equiv \sum_{i\in\Gamma} d(X_i)^2, \]
where the $X_i, i\in\Gamma$ are representers for these classes. If $\2C$ is braided 
then there is another numerical invariant, which we call the Gauss sum, defined by
\[ \Delta_\2C \equiv \sum_{i\in\Gamma} d(X_i)^2 \omega(X_i)^{-1}. \]
The dimensions in a T$C^*$ (not necessarily braided!) are quantized \cite{lro} in the same
way as the square roots of indices in subfactor theory:
\[ d(\rho)\in \left\{ 2\cos\frac{\pi}{n}, n\ge 3 \right\} \cup [2, \infty). \]
The twist $\omega(\rho)$ of a simple object may a priori take any value in the circle
group $\7T$. In a finite dimensional T$C^*$, every $d(\rho)$ is a totally real algebraic
integer and $\omega(\rho)$ is a root of unity. In the braided case there is no known
replacement for Thm.\ \ref{DR-1}.

The deviation of a braided category $\2C$ from being symmetric is measured by the
monodromies 
\[ c_M(X,Y)=c(Y,X)\circ c(X,Y)\in\Hom(XY,XY),
  \quad X, Y\in\obj\,\2C. \]
If $\2C$ has conjugates (in the sense of Thm.\ \ref{thm1}) and the unit $\11$ is simple then
\[ S'(X,Y)\id_\11= (r_X^*\otimes \ol{r}^*_Y) \circ (id_{\ol{X}}\otimes
  c_M(X,Y)\otimes \id_{\ol{Y}})  \circ (r_X\otimes \ol{r}_Y) \]
defines a number which depends only on the isomorphism classes of $X, Y$.
These numbers, for irreducible $X, Y$, were called statistics characters in
\cite{khr1}. (They also give the invariant for the Hopf link with the two components
colored by $X, Y$.) Picking arbitrary representers $X_i, i\in\Gamma$ we define the matrix  
$S'_{i,j}=S'(X_i, X_j),\ i,j\in\Gamma$. The matrix of statistics characters is of particular
interest if the category is finite dimensional.

Then, as proved independently be Rehren \cite{khr1} and Turaev \cite{t}, if $S'$ is
invertible then 
\[ S=(\dim\2C)^{-1/2} S',\ \ \ T=\left(\frac{\Delta_\2C}{|\Delta_\2C|}\right)^{1/3} Diag(\omega_i)
\]
are unitary and satisfy the relations
\[ S^2=(ST)^3=C, \ \ TC=CT, \]
where $C_{ij}=\delta_{i,\bar{\jmath}}$ is the charge conjugation matrix (which satisfies
$C^2=1$). (Whereas the dimension of a T$C^*$ is always non-zero, this is not true in
general. Yet, when $S'$ is invertible then $\dim\2C\ne 0$, cf.\ \cite{t}.)
Since these relations give a presentation of the modular group $SL(2,\7Z)$
we obtain a finite dimensional unitary representation of the latter, which motivated
the terminology `modular category' \cite{t}. Furthermore, the `fusion coefficients'
$N_{ij}^k=\dim\Hom(X_iX_j,X_k)$ are given by the Verlinde relation \cite{v}
\[ N_{ij}^k=\sum_m \frac{S_{im}S_{jm}S_{km}^*}{S_{0m}} .\]

The assumption that $S'$ is invertible is not very conceptual and therefore unsatisfactory.
A better understanding of its significance is obtained from the following considerations.

\bdefin Let $\2C$ be a braided monoidal category and $\2K$ a full subcategory. Then the
relative commutant $\2C\cap\2K'$ of $\2K$ in $\2C$ is the full subcategory defined by 
\[ \obj \ \2C\cap\2K'=\{X\in\2C\ | \  c_M(X,Y)=\id_{XY}\quad \forall Y\in\2K \}. \]
($\2C\cap\2K'$ is automatically monoidal and replete.) The center of a braided monoidal
category $\2C$ is $\2Z(\2C)=\2C\cap\2C'$. 
\label{commut}\edefin
\brem
1. If there is no danger of confusion about the ambient category $\2C$ we will
occasionally write $\2K'$ instead of $\2C\cap\2K'$. 

2. $\2Z(\2C)$ is a symmetric tensor category for every $\2C$. $\2C$ is symmetric iff
$\2Z(\2C)=\2C$. 

3. The objects of the center have previously been called {\it degenerate} 
(Rehren), {\it transparent} (Brugui\`{e}res) and {\it pseudotrivial} (Sawin).
Yet, calling them {\it central} seems the best motivated terminology since the above
definition is the correct analogue for braided tensor categories of the center of a
monoid, as can be seen appealing to the theory of $n$-categories.

4. We say a semisimple category (thus in particular a BT$C^*$) has trivial center, denoted
symbolically $\2Z(\2C)=\11$, if every object of $\2Z(\2C)$ is a direct sum of copies of
the monoidal unit $\11$ of if, equivalently, every simple object in $\2Z(\2C)$ is
isomorphic to $\11$. 

5. Note that the center of a braided tensor category as given in Defin.\ \ref{commut} must 
not be confused with another notion of center \cite{js2,maj} which is defined for all
tensor categories (not necessarily braided) and which in a sense generalizes the quantum
double of Hopf algebras. See also Subsect.\ \ref{cat-4}.
\erem

\bprop\cite{khr1} Let $\2C$ be a BT$C^*$ with finitely many classes of simple
objects. Then the following are equivalent:
\begin{itemize}
\item[(i)] The $S'$-matrix is invertible, thus $\2C$ is modular.
\item[(ii)] The center of $\2C$ is trivial.
\item[(iii)] $|\Delta_\2C|^2=\dim\2C$.
\end{itemize}
\label{crit}\eprop

\brem The direction $(i)\impl(ii)$ is obvious, and $(ii)\impl(i)$ has been generalized by
Brugui\`{e}res \cite{brug3} to a class of categories without $*$-operation, in fact over
arbitrary fields. He proves that a `pre-modular' category \cite{brug1} is modular iff its
dimension is non-zero (which is automatic for $*$-categories) and its center is trivial.
This provides a very satisfactory characterization of modular categories and we see that
modular categories are related to symmetric categories like factors to commutative von
Neumann algebras. Recalling that finite dimensional {\it symmetric} BT$C^*$s are
representation categories of finite groups by the DR duality theorem, one might say that
modular categories ($\2Z(\2C)=\11$) differ from finite groups ($\2Z(\2C)=\2C$) by the
change of a single symbol in the respective definitions!
\erem

\subsection{General Low Dimensional Superselection Theory}
As already mentioned, in low dimensions the category $DHR(\2A)$ is only braided. As a
consequence the proofs \cite{khr1, gl1} of the existence of conjugate (dual)
representations have to proceed in a fashion completely different from \cite[II]{dhr3}.
More importantly, Thm.\ \ref{DR-1} and, a fortiori, Thm.\ \ref{DR-2} are no more
applicable. (There is a weak substitute for the DR field net, cf.\ \cite{frs2, gl1} for
the reduced field bundle, which however is not very useful in practice.) The facts
expounded in the Categorical Interlude imply that every low dimensional QFT whose DHR
category has finitely many simple objects and trivial center gives rise to a unitary
representation of $SL(2,\7Z)$. This is consistent with the physics literature on rational
conformal models but at first sight rather surprising in non-conformal models. (Note,
however, that Haag dual theories which are massive in a certain strong sense have trivial
DHR representation theory \cite{mue3}, implying that for them the question concerning the
r\^{o}le of $SL(2,\7Z)$ does not arise.)

What remains is the issue of triviality of the center of $DHR_f(\2A)$ which does not
obviously follow from the axioms. A first result in this direction was the following which
proves a conjecture in \cite{khr1}. 
\btheor \cite{mue4} Let $\2A$ be a Haag dual theory on $1+1$ dimensional Minkowski space 
or on $\7R$. Assume that $DHR_f(\2A)$ is finite and that all objects in $\2Z(DHR_f(\2A))$
are even, i.e. bosonic. Then $\2F=\2A\rtimes\2Z(DHR_f(\2A))$ is local and Haag dual and 
$DHR_f(\2F)$ has trivial center, thus is modular.
\label{modularization_0}\etheor

In other terms, every rational QFT whose representation category has non-trivial center is
the fixpoint theory of a theory with modular representation category under the action of a
finite group of global symmetries. In the next subsection we will cite results according
to which a large class of models automatically has a modular representation category.
For these models the above theorem is empty, but the analysis of \cite{mue4} is still
relevant for the study of $\2F=\2A\rtimes\2S$ where $\2S\subset DHR_f(\2A)$ is any full
symmetric subcategory, not necessarily contained in $\2Z(DHR_f(\2A))$.

\subsection{Completely Rational Chiral Conformal Field Theories}
In this section we consider chiral conformal field theories, i.e.\ quantum field theories
on the circle. We refer to \cite{mue13} for a more complete and fairly self-contained
account. Let $\2I$ be the set of intervals on $S^1$, i.e.\ connected open non-dense
subsets of $S^1$. For every $J\subset S^1$, $J'$ is the interior of the complement of
$J$, and for $M\subset\2B(\2H_0)$ we posit $M'=\{ x\in\2B(\2H_0)\ | \ xy=yx\ \forall y\in M\}$.
\bdefin A chiral conformal field theory consists of 
\begin{enumerate}
\item A Hilbert space $\2H_0$ with a distinguished non-zero vector $\Omega$,
\item an assignment $\2I\ni I\mapsto \2A(I)\subset\2B(\2H_0)$ of von Neumann algebras to
intervals,
\item a strongly continuous unitary representation $U$ on $\2H_0$ of the M\"obius group 
$PSU(1,1)=SU(1,1)/\{\11,-\11\}$, i.e.\ the group of those fractional linear maps
$\7C\rarr\7C$ which map the circle into itself.
\end{enumerate}
These data must satisfy
\begin{itemize}
\item Isotony: $\ I\subset J\ \ \impl \ \2A(I)\subset\2A(J)$,
\item Locality: $I\subset J'\ \impl \ \2A(I)\subset\2A(J)'$,
\item Irreducibility: $\vee_{I\subset S^1}\2A(I)=\2B(\2H_0)\ $ (equivalently,
$\cap_{I\in\2I} \2A(I)'=\7C\11$),
\item Covariance: $U(a)\2A(I)U(a)^*=\2A(aI)\quad\forall a\in PSU(1,1),\ I\in\2I$,
\item Positive energy: $L_0\ge 0$, where $L_0$ is the generator of the rotation subgroup
of $PSU(1,1)$,
\item Vacuum: every vector in $\2H_0$ which is invariant under the action of $PSU(1,1)$ is a
multiple of $\Omega$.
\end{itemize}
\edefin

For consequences of these axioms see, e.g., \cite{gafr}. We limit ourselves to pointing
out some facts:
\begin{itemize}
\item Reeh-Schlieder property: $\ol{\2A(I)\Omega}=\ol{\2A(I)'\Omega}=\2H_0 \  \ \forall I\in\2I$.
\item Type: The von Neumann algebra $\2A(I)$ is a factor of type $III_1$ for every $I\in\2I$.
\item Haag duality: $\2A(I)'=\2A(I')\ \  \forall I\in\2I$.
\item The modular groups and conjugations associated with $(\2A(I),\Omega)$ have a
geometric meaning, cf.\ \cite{bgl, gafr}.
\end{itemize}

Now one studies coherent representations $\pi=\{\pi_I, I\in\2I \}$ of $\2A$ on
Hilbert spaces $\2H$, where $\pi_I$ is a representation of $\2A(I)$ on $\2H$ such that 
\[ I\subset J \quad \impl \quad \pi_J\restr \2A(I)=\pi_I. \]
One can construct \cite{frs2} a unital $C^*$-algebra $C^*(\2A)$, the global algebra of
$\2A$, such that the coherent representations of $\2A$ are in one-to-one correspondence
with the representations of $C^*(\2A)$. We therefore simply speak of representations.
A representation is covariant if there is a positive energy representation $U_\pi$ of the
universal covering group $\widehat{PSU(1,1)}$ of the M\"obius group on $\2H$ such that  
\[ U_\pi(a)\pi_I(x)U_\pi(a)^* = \pi_{aI}(U(a)xU(a)^*) \quad \forall
   a\in\widehat{PSU(1,1)}, I\in\2I. \]
A representation is locally normal iff each $\pi_I$ is strongly continuous. 

In order to obtain further results we introduce additional axioms.
\bdefin Two disjoint intervals $I, J\in\2I$ are called adjacent if they have exactly one
common boundary point. A chiral CQFT satisfies strong additivity if 
\[ I, J\  \mbox{adjacent} \quad\impl\quad \2A(I)\vee\2A(J)=\2A((I\cup J)''). \]
A chiral CQFT satisfies the split property if $I, J\in\2I$ such that
$\ol{I}\cap\ol{J}=\emptyset$ implies the existence of an isomorphism 
\[ \eta:\ \2A(I)\vee\2A(J)\rarr \2A(I)\otimes\2A(J) \]
of von Neumann algebras satisfying
$\eta(xy)=x\otimes y\ \forall x\in\2A(I), y\in\2A(J)$. 
\edefin

\brem By M\"obius covariance strong additivity holds in general if it holds for one pair
$I, J$ of adjacent intervals. Strong additivity has been verified in all known rational
models. Furthermore, every CQFT can be extended canonically to one satisfying strong
additivity. If the split property holds then $\2H_0$ is separable, and thanks to the
Reeh-Schlieder theorem $\2A(I)\vee\2A(J)$ and $\2A(I)\otimes\2A(J)$ are actually unitarily
equivalent. The split property follows if $Tr e^{-\beta L_0}<\infty$ for all $\beta>0$,
which is satisfied in all reasonable models. 
\erem

\blemma\cite{klm}
Let $\2A$ be a CQFT satisfying strong additivity and the split property. Let 
$I_k\in\2I, k=1,\ldots,n$ be intervals with mutually disjoint closures and denote 
$E=\cup_k I_k$. Then $\2A(E)\subset\2A(E')'$ is an irreducible inclusion (of type $III_1$
factors) and the index $[\2A(E')':\2A(E)]$ depends only on the number $n$ but not on the
choice of the intervals. Let $\mu_n$ be the index for the $n$-interval inclusion. These
numbers are related by 
\[ \mu_n=\mu_2^{n-1} \quad\forall n\in\7N. \]
\elemma
(In particular $\mu_1=1$, which is just Haag duality.) Thus every CQFT satisfying strong
additivity and the split property comes along with a numerical invariant $\mu_2\in
[1,\infty]$ whose meaning is elucidated by the main result of \cite{klm} stated below.

\bdefin A chiral CQFT is completely rational if it satisfies (a) strong additivity, (b)
the split property and (c) $\mu_2<\infty$.
\edefin

All known classes of rational CQFTs are completely rational in the above sense, see
\cite{wass,x1} for the WZW models connected to loop groups and \cite{x2, mue11} for
orbi\-fold models. Very strong results on both the structure and representation theory 
of completely rational theories can be proved. (All representations are understood to be
non-degenerate.) 

\btheor \cite{klm, mue13} \label{KLM_main} 
Let $\2A$ be a completely rational CQFT. Then
\begin{itemize}
\item Every representation of $C^*(\2A)$ on a separable Hilbert space is locally normal
and completely reducible, i.e.\ the direct sum of irreducible representations.
\item Every irreducible separable representation has finite statistical dimension
$d_\pi\equiv[\pi(\2A(I'))':\pi(\2A(I))]^{1/2}$ (independent of $I\in\2I$).
It therefore \cite{gl1} has a conjugate representation $\ol{\pi}$ and is automatically
M\"obius covariant with positive energy. 
\item For a representation $\pi$ the following are equivalent: (a) $\pi$ is M\"obius
covariant with positive energy, (b) $\pi$ is locally normal, (c) $\pi$ is a direct sum of
separable representations.
\item The category $Rep(\2A)$ of finite direct sums of separable irreducible
representations has a monoidal structure with simple unit and duals (conjugates).
The number of unitary equivalence classes of separable irreducible representations
is finite and 
\[ \dim Rep(\2A)=\mu_2(\2A). \]
Furthermore, there is a non-degenerate braiding. $Rep(\2A)$ thus is a unitary modular
category in the sense of Turaev \cite{t}. 
\end{itemize}
\etheor

\brem 1. In the way of structure theoretical results we mention that for completely rational
theories the subfactors $\2A(E)\subset\2A(E')'$, $E=\cup_{i=1}^n I_i$ can be analyzed
quite explicitly, generalizing some of the results of \cite{x1}. Yet \cite{klm} by no
means supersedes the ingenious computation in \cite{x1} of the indices $[\2A(E')':\2A(E)]$
in the case of loop group models.  

2. In view of the above results we do not need to worry about representations with
infinite statistics when dealing with completely rational CQFTs. From now on we will write
$Rep(\2A)$ instead of $DHR(\2A)$ since the (separable) representation theory can be
developed without any selection criterion \cite{mue13}. Some of our results hold for
low-dimensional theories without the assumption of complete rationality. For this we refer
to \cite{mue11}.
%
\erem

\section{From $Rep(\2A)$ to $Rep(\2F)$}
\subsection{Pairs of Quantum Field Theories Related by a Symmetry Group}
In the rest of this paper we will be concerned with pairs $(\2F, \2A)$ of quantum field
theories in one or two dimensions where $\2F$ has a compact group $G$ of global symmetries
(acting non-trivially for $g\ne e$) and $\2A=\2F^G\restr\2H_0$. We assume that both $\2A$
and $\2F$ satisfy Haag duality. Then there is a full symmetric subcategory $\2S\subset
Rep(\2A)$ such that $\2S\simeq G-\mod$ and $\2F\cong\2A\rtimes\2S$. This situation is
summarized in the quadruple $(\2F, G; \2A, \2S)$. Our aim will be to compute the
representation category of $\2F$ from that of $\2A$ and vice versa. The nicest case
clearly is the one where both $\2A$ and $\2F$ are completely rational CQFTs (then $G$ must
be finite), but some of our results hold in larger generality. 

\btheor\cite{x2, klm, mue11}\label{A_comp_ratl}
Consider a pair $(\2F, G; \2A, \2S)$ of chiral theories. If $\2A$ is completely rational
then $\2F$ is completely rational. With $\2F$ completely rational, $\2A$ is completely
rational iff $G$ is finite. In this case $\mu_2(\2A)=|G|^2 \mu_2(\2F)$, thus
\begin{equation}
 \dim Rep(\2A)=|G|^2 \dim Rep(\2F). 
\label{dima}\end{equation}
\etheor

\brem That fixpoint nets inherit the split property from field nets is classical
\cite{dopl}, and that $\2F$ satisfies strong additivity if $\2A$ does is almost
trivial. The converses of these two implications are non-trivial and require the full
force of complete rationality. The implication $\2F$ completely rational $\impl\ \2A$
satisfies strong additivity is proved in \cite{x2}, and $\2A$ completely rational 
$\impl\ \2F$ satisfies split will be proved below. The computation of the invariant
$\mu_2(\2A)$ is done already in \cite{klm}.
\erem

\brem The completely different structure (symmetric instead of modular) of the
representation categories in $\ge 2+1$ dimensions is reflected in a replacement of $|G|^2$
in (\ref{dima}) by $|G|$. 
\erem

In Subsect.\ \ref{ss:DHRF} we will show that the representation category of $\2F$ depends
only on $Rep(\2A)$ and the symmetric subcategory $\2S$. More precisely, let $\2A_1, \2A_2$
be QFTs such that $Rep(\2A_1)\simeq Rep(\2A_2)$ and let $\2S_i\subset Rep(\2A_i), i=1,2$
be replete full symmetric subcategories which correspond to each other under the above
equivalence. Then we claim $Rep(\2F_1)\simeq Rep(\2F_2)$ where
$\2F_i=\2A_i\rtimes\2S_i$. The most natural way to prove such a result clearly is to
construct a braided tensor category from $Rep(\2A)$ and $\2S$ and to prove that it is
equivalent to $Rep(\2A\rtimes\2S)$ independently of the fine structure of $\2A$. The next
categorical interlude will provide such a construction.

\subsection{Categorical Interlude 2: Galois Extensions of Braided Tensor Categories}\label{Int-2}
The following result realizes a conjecture in \cite{mue4}.
\btheor \cite{brug1, mue6}
Let $\2C$ be a BT$C^*$. Let $\2S\subset\2C$ be a replete full monoidal subcategory which
is symmetric (with the braiding of $\2C$). Then there exists a T$C^*\ \2C\rtimes\2S$
together with a tensor functor $F: \2C\rarr\2C\rtimes\2S$ such that 
\begin{itemize}
\item $F$ is faithful and injective on the objects, thus an embedding.
\item $F$ is dominant, i.e.\ for every simple object $X\in\2C\rtimes\2S$ there is
$Y\in\2C$ such that $X$ is a subobject of $F(Y)$.
\item $F$ trivializes $\2S$, i.e.\ $X\in\2S\ \impl\ F(X)\cong \11 \oplus \ldots \oplus \11$, 
where $\11$ appears with multiplicity $d(X)$ (which is in $\7N$ by \cite{dr6}).
\item The pair $(\2C\rtimes\2S, F)$ is the universal solution for the above problem, i.e.\
if $F': \2C\rarr \2E$ has the same properties then $F'$ factorizes through $F$.
\end{itemize}
\etheor

\brem This result was arrived at independently by the author \cite{mue6} and (somewhat
earlier) by Brugui\`{e}res \cite{brug1}. The above statement incorporates some results of
\cite{brug1}. The construction in \cite{brug1} relying on Deligne's duality theorem
\cite{del} instead of the one of \cite{dr2} it is slightly more general, but one must
assume that the objects in $\2S$ have integer dimension since this is no more automatic if
there is no positivity. On the other hand, in \cite{brug1} $\2S$ is assumed finite
dimensional (thus $G$ is finite) and to be contained in $\2Z(\2C)$, restrictions which are
absent in \cite{mue6}. Applications of the above construction to quantum groups and
invariants of 3-manifolds are found in \cite{brug1} and \cite{saw}, the latter reference
considering also relations with products of braided categories and of TQFTs.
\erem

\brem By the universal property $\2C\rtimes\2S$ is unique up to equivalence. The existence
is proved by explicit construction. Essentially, one adds morphisms to $\2C$ which
trivialize the objects in $\2S$. (Then one completes such that all idempotents split, but
this is of minor importance.) Here essential use is made of the fact that there is a
compact (respective finite) group $G$ such that $\2S\simeq G-\mod$. 
\erem

Many facts are known about the category $\2C\rtimes\2S$:

\bprop\cite{brug1, mue10}\label{dim_CS}
If $\2C$ is finite dimensional then
\begin{equation}
  \dim \2C\rtimes\2S= \frac{\dim\2C}{\dim\2S}=\frac{\dim\2C}{|G|}.
\label{dimcs}\end{equation}
\eprop

\brem Heuristically, the passage from $\2C$ to $\2C\rtimes\2S$ amounts to dividing out the
subcategory $\2S$, an idea which is further supported by (\ref{dimcs}). Yet, this is not
done by killing the objects of $\2S$ in a quotient operation but rather by adding
morphisms which trivialize them. Therefore the notation $\2C\rtimes\2S$, which is also in
line with \cite{dr2}, seems more appropriate. We consider $\2C\rtimes\2S$ as a Galois
extension of $\2C$ as is amply justified by the following result.
\erem 

\bprop \cite{mue6,brug2} We have $G\cong\mbox{Aut}_\2C(\2C\rtimes\2S)$ and there is a
Galois correspondence between closed subgroups $H$ of $G$ and T$C^*$s $\2E$
satisfying $\2C\subset\2E\subset\2C\rtimes\2S$. (The correspondence is given by 
$\2E=(\2C\rtimes\2S)^H$ and $H=\mbox{Aut}_\2E(\2C\rtimes\2S)$.)
Here $H$ is normal iff $\2E=\2C\rtimes\2T$ with $\2T$ a replete full subcategory of $\2S$,
in which case $\mbox{Aut}_\2C(\2E)\cong G/H$.
\eprop

\btheor \cite{mue6} \label{modularization}
The braiding of $\2C$ lifts to a braiding of $\2C\rtimes\2S$ iff $\2S\subset\2Z(\2C)$. In
this case $\2C\rtimes\2S$ has trivial center iff $\2S=\2Z(\2C)$. $\2C\rtimes\2Z(\2C)$ is
called the modular closure $\ol{\2C}^m$ of $\2C$ since it is modular if $\2C$ is finite
dimensional.
\etheor

\brem This result has obvious applications to the topology of 3-mani\-folds since it
provides a means of constructing a modular category out of every finite dimensional
braided tensor category (which must not be symmetric). In fact, ad hoc versions of the
above constructions in simple special cases motivated by topology had appeared before.
\erem

If $\2S\not\subset\2Z(\2C)$ then $\2C\rtimes\2S$ fails to have a braiding in the usual
sense. Yet, there is a braiding in the following generalized sense.

\bdefin \label{graded1}
Let $\2C$ be a semisimple $k$-linear category over a field $k$. If $G$ is a group
then $\2C$ is $G$-graded if
\begin{enumerate}
\item With every simple object $X$ is associated an element $gr(X)\in G$. 
\item If $X, Y$ are simple and isomorphic then $gr(X)=gr(Y)$.
\item Let $\2C_g$ be the full subcategory of $\2C$ whose objects are finite direct sums of
objects with grade $g$. Then $X\in\2C_g, Y\in\2C_h$ implies $X\otimes Y\in\2C_{gh}$. 
\end{enumerate}
If $\2C$ is $G$-graded and carries a $G$-action such that
\[ \alpha_g (\2C_h) = \2C_{ghg^{-1}} \quad \forall g,h\in G, \]
then $\2C$ is a crossed $G$-category. A crossed $G$-category is braided if there are
isomorphisms $c(X,Y): X\otimes Y\rarr \alpha_{gr(X)}(Y)\otimes X$ for all $Y$ and all
homogeneous $X$, i.e.\ $X\in\2C_g$ for some $g\in G$. For the relations which $c$ must
satisfy cf.\ \cite{t2}.
\edefin

\brem Our definitions are slightly more general than those given by Turaev in \cite{t2}
in that we allow for direct sums of objects of different grade. This complicates the
definition of the braiding, but the gained generality is needed in our applications. 
\erem

\btheor \cite{mue10} \label{graded2}
Let $\2C$ be a BT$C^*$ and $\2S$ a replete full symmetric subcategory. Then
$\2C\rtimes\2S$ is a crossed $G$-category. The zero-grade part of $\2C\rtimes\2S$ is 
given by
\[ (\2C\rtimes\2S)_e = (\2C\cap\2S')\rtimes \2S, \]
which is always a BT$C^*$ and has trivial center iff $\2Z(\2C)\subset\2S$. 
The set $H$ of $g\in G$ for which $\2C_g$ is non-empty is a closed normal subgroup which
corresponds to $\2S\cap\2Z(\2C)$ under the bijection between closed normal subgroups of
$G$ and replete full monoidal subcategories of $\2S$. Thus the grading is full 
($(\2C\rtimes\2S)_g\ne\emptyset\ \forall g\in G$) iff $\2S\cap\2Z(\2C)=\11$ and trivial 
($\2C\rtimes\2S=(\2C\rtimes\2S)_e$) iff $\2S\subset\2Z(\2C)$. If $\2C$ is modular then
$\2C\rtimes\2S$ is modular in the sense of \cite{t2} with full grading for every $\2S$.
\etheor

This result will be relevant when we compute $Rep(\2A)$ in Sect.\ \ref{DHR-A}.

\bprop\cite{mue6} Let $X\in\2C$ be simple. Then all simple subobjects $X_i$ of
$F(X)\in\2C\rtimes\2S$ occur with the same multiplicity and have the same dimension. If
$\2S\subset\2Z(\2C)$, thus $\2C\rtimes\2S$ is braided, then all $X_i$ have the same twist
as $X$, and they are either all central or all non-central according to whether $X$ is
central or non-central. 
\label{crucial}\eprop 

Given irreducible objects $X\not\cong Y$ in $\2C$ we should also understand whether
they can have equivalent subobjects in $\2C\rtimes\2S$. We have

\bprop \cite{mue10} Let $X, Y$ be simple objects in $\2C$. We write $X\sim Y$ iff there is
$Z\in\2S$ such that $\Hom(ZX,Y)\ne\{0\}$. This defines an equivalence relation which is
weaker than isomorphism $X\cong Y$. If $X\sim Y$ then $F(X)$ and $F(Y)$ contain the same
(isomorphism classes of) simple objects of $\2C\rtimes\2S$ (whose multiplicity in $F(X)$
and $F(Y)$ need not be the same), otherwise $\Hom(F(X),F(Y))=\{0\}$.
\eprop

In the case of abelian extensions one can give a more complete analysis. Recall that $G$
is abelian iff every simple object in $\2S\cong G-\mod$ has dimension one (equivalently,
is invertible up to isomorphism). In this case the set of isomorphism classes of simple
objects in $\2S$ is an abelian group $K\cong\hat{G}$ (as opposed to an abelian semigroup
in the general case). Since the tensor product of a simple and an invertible object is
simple, $K$ acts on the set $\Gamma$ of isomorphism classes of simple objects of $\2C$ (by
tensoring of representers). For every simple $X\in\2C$ we define 
$K_X=\{[Z]\in K\ | \ [Z][X]=[X]\}$, the stabilizer of $X$, which is a finite subgroup of
$K$. (Non-trivial stabilizers can exist since generically there is no cancellation in
$\Gamma$!) 

\bprop \cite{mue6} \label{sce1}
For every simple $X\in\2C$ there is a subgroup $L_X\subset K_X$ such that
$N_X=[K_X:L_X]^{1/2}\in\7N$ and
\[ F(X)\cong N_X \bigoplus_{\chi\in\widehat{L_X}} X_\chi, \]
where the $X_\chi$ are mutually inequivalent simple objects in $\2C\rtimes\2S$.
$K_X$ and $L_X$ depend only on the image $\ul{X}$ of $X$ in 
$\Gamma/K$ (i.e.\ the $K$-orbit in $\Gamma$ which contains $[X]$). The isomorphism
classes of simple objects in $\2C\rtimes\2S$ are labeled by pairs $(\ul{X}, \chi)$, where
$\ul{X}\in\Gamma/K$ and $\chi\in\widehat{L_{\ul{X}}}$.
\eprop

\bprop \cite{mue10} \label{sce2} Assume $\2S\subset\2Z(\2C)$. (If necessary enforce this
by replacing $\2C\rarr\2C\cap\2S'$.) There are unitary matrices
$(S^{[Z]}(\ul{X},\ul{Y}))_{\ul{X},\ul{Y}}$ where $\ul{X},\ul{Y}\in\Gamma/K$ are
$Z$-fixpoints (i.e.\ $[Z]\in K_{\ul{X}}\cap K_{\ul{Y}}$), such that
\begin{equation} 
S_{\2C\rtimes\2S}((\ul{X},\chi),(\ul{Y},\nu)) =
  \frac{|G|}{|K_{\ul{X}}||L_{\ul{X}}||K_{\ul{Y}}||L_{\ul{Y}}|}
  \sum_{Z\in L_{\ul{X}}\cap L_{\ul{Y}}} \chi(Z)\ol{\nu(Z)}\, S^{[Z]}(\ul{X},\ul{Y}).
\label{S-F}\end{equation}
Together with the (appropriately restricted) $T$-matrix of $\2C$ the matrices $S^{[Z]}$
satisfy the relations of the mapping class group of the torus with a puncture \cite{bv}.
\eprop

The two preceding propositions are abstract versions of heuristically derived results in
\cite{fss} which provided decisive motivation.

\subsection{Computation of $Rep(\2F)$}\label{ss:DHRF}
Given a $\rho\in Rep(\2A)$ of $\2A$ which is localized in some double cone $\2O$ there
exists \cite{khr2,mue4} an extension $\hat{\rho}$ to an endomorphism of
$\2F=\2A\rtimes\2S$ which commutes with the action of $G$. It is determined by
\[ \hat{\rho}(\psi)=c(\gamma,\rho)\psi, \quad \mbox{where} \quad \psi\in\2H_\gamma, \]
where $\gamma\in\2S$ and the spaces
\[ \2H_\gamma=\{ \psi\in\2F \ | \ \psi A=\gamma(A)\psi \ \ \forall A\in\2A \} \]
generate $\2F$ linearly. In $\ge 2+1$ dimensions this extension is unique and again
localized in $\2O$ since $c(\gamma,\rho)=1$ whenever $\rho, \gamma$ have spacelike
localization regions. For theories in $1+1$ dimensional Minkowski space or on $\7R$,
however, there is an a priori different extension obtained by replacing $c(\gamma,\rho)$
by $c(\rho,\gamma)^*$, and for spacelike localized $\rho, \gamma$ a priori only one of
these equals $\11$. Thus the two extensions are solitonic, i.e.\ localized in left and
right, respectively, wedges or half-lines. They coincide and are localized in $\2O$ iff
the two braidings are the same for all $\gamma\in\2S$, thus precisely if $\rho\in
Rep(\2A)\cap\2S'$. For theories on $S^1$ an extension $\hat{\rho}$ does not even exist as
a representation of $C^*(\2A)$ if $\rho\not\in\2S'$. The map  
$Rep(\2A)\cap\2S'\rarr Rep(\2F), \ \rho\mapsto\hat{\rho}$ being functorial, it follows 
easily from the definition of $\2F=\2A\rtimes\2S$ and $\2C\rtimes\2S$ that 
$(Rep(\2A)\cap\2S')\rtimes \2S$ 
is (equivalent as a braided monoidal category to) a replete full monoidal subcategory of 
$Rep(\2F)$. In fact, by an argument similar to the one used in Sect.\ \ref{doro} 
one can prove that this exhausts the sectors of $\2F$ and one obtains:

\btheor \cite{mue11} \label{AtoF}
The representation category of $\2F$ is given (up to equivalence of
braided tensor $*$-categories) by
\begin{equation}
   Rep(\2F) \simeq (Rep(\2A)\cap \2S')\rtimes \2S.
\label{dhrf}\end{equation}
Thus $Rep(\2F)$ depends only on $Rep(\2A)$ and the symmetric subcategory $\2S$. 
\etheor
(This result together with Thm.\ \ref{modularization} provides another proof of Thm.\ 
\ref{modularization_0}.)

The operation $\2A\rarr\2F=\2A\rtimes\2S$ where $G$ is abelian is called a {\it simple
current extension}. Bringing to bear our results on abelian Galois extensions we obtain 
the following theorem which proves most of the observations of \cite{fss} many of which
were based on consistency checks rather than proofs.

\btheor \cite{mue11} Consider $(\2F,G; \2A,\2S)$ with $G$ abelian. Then the equivalence
classes of irreducible localized representations of $\2F$ are labeled by pairs $(\ul{X},
\chi)$, where $\ul{X}\in\obj(Rep(\2A)\cap\2S')/\cong/K$ and
$\chi\in\widehat{L_{\ul{X}}}$. Here $K\cong\hat{G}$ is the set of isomorphism classes of
simple objects in $\2S$ and $L_{\ul{X}}$ is as in Prop.\ \ref{sce1}. The modular
$S$-matrix of $Rep(\2F)$ is given by formula (\ref{S-F}).
\etheor

Contrary to the fixpoint problem $\2F\rarr\2A=\2F^G$, non-abelian {\it extensions}
$\2A\rarr\2F=\2A\rtimes\2S$ seem not to have been considered in the physics
literature. (This is perhaps not surprising since they require the duality theorems
either of Doplicher/Roberts or Deligne.)

Assuming that $\2A$ is completely rational we know by Thm.\ \ref{A_comp_ratl} that $\2F$
is completely rational, thus $Rep(\2F)$ is modular. In view of the `explicit' formula
(\ref{dhrf}) for $Rep(\2F)$ and of Thm.\ \ref{modularization} we can conclude that
\begin{equation} 
   \2Z(Rep(\2A)\cap \2S')=\2S. 
\label{Z1}\end{equation}
Furthermore, the dimension of $Rep(\2F)$ is given by $\dim Rep(\2A)/|G|^2$. 
Comparing this with (\ref{dimcs}) we infer
\begin{equation}
   \dim (Rep(\2A)\cap \2S')=\frac{\dim Rep(\2A)}{\dim\2S}. 
\label{Z2}\end{equation}
There should clearly be a purely categorical proof of these two observations. In fact, the
result holds in considerably larger generality and is the subject of our next categorical
interlude.

\subsection{Categorical Interlude 3: Double Commutants in Modular Categories}
For obvious reasons the following result will be called the (double) commutant theorem for
modular categories. 
\btheor\cite{mue10} Let $\2C$ be a modular BT$C^*$ and let $\2K\subset\2C$ be a replete full
sub T$C^*$ closed w.r.t.\ direct sums and subobjects (as far as they exist in $\2C$). 
Then we have 
\begin{itemize}
\item[(a)] $\displaystyle \2K''=\2K$,
\item[(b)] $\displaystyle \dim \2K \,\cdot\, \dim\2K'=\dim\2C$.
\end{itemize}
\label{dct}\etheor
\brem The double commutant property (a) appears first (without published proof) in the
notes \cite{ocn3} in connection with Ocneanu's asymptotic subfactor \cite{ocn2}. In
the subfactor setting (a) and (b) are proved in \cite{iz2}. A simple argument proving (a)
and (b) in one stroke in the more general setting of $C^*$-categories appears in
\cite{mue10}. Finally, the theorem was then extended \cite{brug3} to categories $\2C$
which are semisimple spherical with non-zero dimension. It seems likely that this is the
most general setting where it holds.
\erem

Thm.\ \ref{dct} has many applications, the first of which is the desired purely
categorical proof of (\ref{Z1}). Let thus $\2C$ be modular and $\2S$ a replete full
monoidal subcategory. Then 
\begin{equation}
\2Z(\2C\cap\2S')=\2C\cap\2S'\cap(\2C\cap(\2C\cap\2S')')=\2S'\cap\2S''=\2S'\cap\2S=\2Z(\2S).
  \label{cent}\end{equation}
If $\2S$ is symmetric, thus $\2Z(\2S)=\2S$, (\ref{Z1}) follows at once, and (\ref{Z2})
is just a special case of (b).

Consider now a modular category $\2C$ with a replete full {\it modular} subcategory
$\2K$. Modularity being equivalent to triviality of the center by Prop.\ \ref{crit},
(\ref{cent}) implies the following.
\bcoro Let $\2C$ be a modular BT$C^*$ and let $\2K\subset\2C$ be a replete full modular 
sub T$C^*$. Then $\2L=\2C\cap\2K'$ is modular, too. \ecoro

A surprisingly easy argument now proves:
\btheor\cite{mue10} Let $\2K\subset\2C$ be a replete full inclusion of modular BT$C^*$s
and let $\2L=\2C\cap\2K'$. Then there is an equivalence of braided monoidal categories:
\[ \2C\simeq\2K\otimes_\7C\2L, \]
where $\otimes_\7C$ is the product in the sense of enriched category theory. 
\label{tprod}\etheor

\brem This result implies that every modular category is a direct product of prime ones,
the latter being defined by the absence of proper replete full modular subcategories. 
Again this holds beyond the setting of $*$-categories \cite{brug3}. The question in which
sense this factorization might be unique is quite non-trivial.

It is also interesting to note the analogy with the well-known result from the theory of
von Neumann algebras where an inclusion $A\subset B$ of type I factors gives rise to an
isomorphism $B\cong A\otimes (B\cap A')$.
\erem

\section{From $Rep(\2F)$ to $Rep(\2A)$} \label{DHR-A}
\subsection{Does $Rep(\2F)$ Determine $Rep(\2A)$?}\label{f_to_a}
By the results of Subsect.\ \ref{ss:DHRF} we have
\[ Rep(\2F) \simeq (Rep(\2A)\cap\2S')\rtimes \2S. \]
The Galois group $G$ (which is determined up to isomorphism by $G-\mod\simeq\2S$) acts on
$Rep(\2F)$ and the fixpoints are given by
\[ Rep(\2F)^G \simeq Rep(\2A)\cap\2S' \subset Rep(\2A). \]
Thus the category $Rep(\2F)^G$, which consists just of those localized transportable
endomorphisms of $\2F$ which commute with all $\alpha_g$, is only a full subcategory of
$Rep(\2A)$, viz. precisely $Rep(\2A)\cap\2S'$. The latter cannot coincide with $Rep(\2A)$
since this would mean $\2S\subset\2Z(Rep(\2A))$, whereas we know that $Rep(\2A)$ has
trivial center.  

Abstractly the situation is the following. We have a non-modular category
$\2C_0=Rep(\2A)\cap\2S'$ and its modular closure 
$\ol{\2C_0}^m=\2C_0\rtimes\2Z(\2C_0)\cong Rep(\2F)$ familiar from Sect.\ \ref{Int-2}.
But we also have a modular category $\2C=Rep(\2A)$ which contains $\2C_0$ as
a full subcategory. The dimensions of the categories in question are:  
\begin{eqnarray}
   \dim \ol{\2C_0}^m &=& \frac{\dim\2C_0}{\dim \2Z(\2C_0)}, \nn \\
   \dim \2C &=& \dim\2C_0 \cdot \dim \2Z(\2C_0). \label{dims}
\end{eqnarray}
This suggests the conjecture that every non-modular category $\2C_0$ embeds as a full
subcategory into a modular category $\2C$ such that (\ref{dims}) holds. We will look into
this problem in the next Categorical Interlude, without however giving a proof.

\subsection{Categorical Interlude 4: Constructing Modular Categories}\label{cat-4}
In Categorical Interlude 2 we have constructed modular categories out of braided
categories by adding morphisms, which heuristically amounts to dividing out the center. 
In Subsection \ref{ss:DHRF} we have seen that this categorical construction reflects what
happens in the passage from $Rep(\2A)$ to $Rep(\2F)$. 

Given a braided tensor category $\2C$ with non-trivial center one might wish to
construct a modular category $\2M$ into which $\2C$ is embedded as a full subcategory,
i.e.\ without tampering with $\2C$ as done in Subsect.\ \ref{Int-2}. 

\blemma Let $\2M$ be a modular BT$C^*$ and $\2C\subset\2M$ a replete full sub T$C^*$. Then
\begin{equation}
   \dim\2M\ge \dim\2C\cdot \dim\2Z(\2C). 
\label{bound}\end{equation}
\elemma
\prf The obvious inclusion
\[ \2M\cap\2C'\supset\2C\cap\2C'=\2Z(\2C) \]
in conjunction with Thm.\ \ref{dct} implies for any modular extension $\2M$
\[ \dim(\2M\cap\2C')=\frac{\dim\2M}{\dim\2C}\ge \dim\2Z(\2C) \]
and thus the bound (\ref{bound}).
\qed

\begin{conjecture} \label{minext}
For every BT$C^*\ \2C$ of finite dimension there exists a modular extension $\2M$ of
dimension
\begin{equation}\label{dimmin}
  \dim\2M=\dim\2C\cdot \dim\2Z(\2C). 
\end{equation}
\end{conjecture}

An equivalent conjecture was formulated, in fact claimed to be true without proof, by
Ocneanu \cite{ocn3}. We do not have any doubt concerning its correctness but,
unfortunately, we are not aware of a proof.
The considerations of the preceding subsection show that the conjecture is in fact true
for all categories of the form $Rep(\2A)\cap\2S'$, where $\2A$ is a CQFT and $\2S$ is a
full  symmetric subcategory of $Rep(\2A)$. Since we do not know that all BT$C^*$s actually
appear as representation categories of some CQFT this provides evidence for the
conjecture, but no proof. 

If $\2C$ is already modular, i.e.\ $\dim\2Z(\2C)=1$, then $\2M=\2C$ clearly is a minimal
modular extension. On the other hand, if $\2Z(\2C)=\2C$ then $\2C\simeq G-\mod$ for a
finite group $G$ and a modular extension of dimension $|G|^2=(\dim\2C)^2$, thus minimal,
is given by $D^\omega(G)-\mod$, where $\omega\in Z^3(G,\7T)$ and $D^\omega(G)$ is the
twisted quantum double \cite{dpr}. Since $D^{\omega_1}(G)-\mod\not\simeq D^{\omega_2}(G)-\mod$  
if $[\omega_1]\ne[\omega_1]$ this example shows already that the minimal extension need
not be unique. But apart from these easy cases it is a priori not obvious that it is at
all possible to fully embed braided tensor categories into modular ones, even in a
non-minimal way. This is proven by the `center construction' for tensor categories
\cite{js2,maj}, a construction which produces a braided tensor category $\2D(\2C)$ out of
any (not necessarily braided!) tensor category $\2C$. If $\2C$ happens to be braided then
it imbeds into $\2D(\2C)$ as a replete full subcategory. The category $\2D(\2C)$
generalizes the quantum double $D(H)$ of a finite dimensional Hopf algebra $H$ in the
sense that there is an equivalence of braided tensor categories
\[ \2D(H-\mod)\simeq D(H)-\mod. \]
(See \cite[Sect.\ XIII.4]{ka} for a nice presentation of all this.)
For this reason -- and also in order to avoid confusion with the center $\2Z(\2C)$ of a
braided tensor category as defined in Sect.\ \ref{Int-1} -- we refer to  $\2D(\2C)$ as the
quantum  double of $\2C$. In \cite{eg} it has been shown $\2D(\2C)$ is a modular category
if $\2C\simeq H-\mod$ where $H$ is a finite dimensional semisimple Hopf algebra over a
field of characteristic zero. This can been generalized to the much wider setting of
tensor categories:
\btheor\cite{mue9} \label{cdouble}
Let $\2C$ be a semisimple spherical tensor category $\2C$ with non-zero
dimension over an algebraically closed field (of arbitrary characteristic). (Finite
dimensional BT$C^*$ belong to this class). Then $\2D(\2C)$ is semisimple, spherical 
and modular with
\begin{equation}\label{dimdc}
 \dim \2D(\2C)=(\dim\2C)^2.
\end{equation}
\etheor

\brem Eq.\ (\ref{dimdc}) clearly is the only identity which is compatible with the
special case $\2C\simeq H-\mod$. Concerning the proofs we must limit ourselves to the remark
that they are based on the adaption \cite{mue8} of results from subfactor theory to
category theory. These works owe much to \cite{iz2} which provides not only the crucial
motivation but also bits of the proof.
\erem

By the above, $\2D(\2C)$ provides a modular extension of $\2C$, which is minimal iff
$\2C=\2Z(\2C)$, i.e.\ if $\2C$ is symmetric, as one sees comparing (\ref{dimdc}) with
(\ref{dimmin}). One might hope that a {\it minimal} modular extension can be constructed
by a modification of the quantum double.

As another application of the double commutant theorem we exhibit a construction which
provides many examples of BT$C^*$s which admit a minimal modular extension. 
\bprop \cite{mue10}
Let $\2C$ be a finite dimensional BT$C^*$ and let $\2E$ be the full monoidal subcategory
of the quantum double $\2D(\2C)$ which is generated by $\2C$ and $\2D(\2C)\cap\2C'$. Then
$\2Z(\2E)=\2Z(\2C)$ and $\dim\2E=(\dim\2C)^2/\dim\2Z(\2C)$, thus 
$\dim\2D(\2C)=\dim\2E\cdot\dim\2Z(\2E)$ and the quantum double $\2D(\2C)$ is a minimal
modular extension of $\2E$.
\eprop

\subsection{Computing $Rep(\2A)$: Soliton Endomorphisms}\label{rep-A}
We have seen that it is not possible to compute $Rep(\2A)$ knowing just $Rep(\2F)$. Thus
we must use properties of $\2F$ which go beyond the localized representations. The aim of
this subsection is to identify the additional information we need. We have already used
the fact that every localized endomorphism $\rho$ of $\2A$ extends to an endomorphism
$\hat{\rho}$ of $\2F$ which is localized iff $\rho\in \2S'$. Thus, trivially, every
$\rho\in Rep(\2A)$ is obtained as restriction of $\hat{\rho}$ to $\2A$. This makes clear
that we should understand the nature of $\hat{\rho}$ for $\rho\not\in\2S'$.

In the following discussion we consider theories on $\7R$ or $1+1$ dimensional Minkowski
space. (In the case of theories living on $S^1$ one must remove an arbitrary point `at
infinity' in order for $\hat{\rho}$ to be well defined.) For any double cone $\2O$ (or
interval $I$) we denote by $\2O_L$ (resp.\ $\2O_R$) its left (resp.\ right) 
spacelike complement. A endomorphism of $\2F$ which acts on $\2F(\2O_R)$ like $\alpha_g$
for some $g\in G$ and as the identity on $\2F(\2O_L)$ is called a right handed $g$-soliton
endomorphism associated with in $\2O$. (Left handed soliton endomorphisms are of course
defined analogously, but it is sufficient to consider one species.) A $G$-soliton
endomorphism is a $g$-soliton associated with some $\2O\in\2K$ for some $g\in G$. We
emphasize that for $\rho\in\2S'$, $\hat{\rho}$ is a bona fide superselection sector
(possibly reducible) of $\2F$, but the soliton endomorphisms $\hat{\rho}$ of the
quasilocal algebra $\2F$ arising if $\rho\not\in\2S'$ provably do not admit extension to
locally normal representations on $S^1$. Heuristically this is clear since they `act
discontinuously at infinity'.

\blemma\cite{mue11} \label{submor}
Consider $(\2F,G;\2A,\2S)$ with $G$ compact but not necessarily finite. Let $\rho$ be an
irreducible transportable endomorphism of $\2A$ localized in a double cone $\2O$. Let
$\hat{\rho}$ be its (right hand localized) extension to $\2F$ and let $\hat{\rho}_1$ be an
irreducible submorphism of $\hat{\rho}$. (If $E\in\2F\cap\hat{\rho}(\2F)'\subset\2F(\2O)$
is the corresponding minimal projection we pick an isometry in $\2F(\2O)$ in order to
define $\hat{\rho}_1$.) Then there is $g\in G$ such that 
\[ \hat{\rho}_1\restr\2F(\2O_R) = \alpha_g. \]
In particular, if $\hat{\rho}$ is irreducible then there is $g\in Z(G)$ such that
\[ \hat{\rho}\restr\2F(\2O_R) = \alpha_g. \]
The latter is, a fortiori, the case if $\rho$ is a localized automorphism of $\2A$.
\elemma

Lemma \ref{submor} shows that every irreducible localized endomorphism of $\2A$ is the
restriction to $\2A$ of a direct sum of $G$-soliton endomorphisms of $\2F$. The following
is more precise.

\bprop \cite{mue11} \label{decomp}
Let $\2A, \2F, \rho, \hat{\rho}$ be as in Lemma \ref{submor}. Then there is a
conjugacy class $c$ of $G$ such that $\hat{\rho}$ contains an irreducible $g$-soliton
endomorphism iff $g\in c$. The adjoint action of the group $G$ on the (equivalence classes
of) irreducible submorphisms of $\hat{\rho}$ is transitive. Thus all irreducible soliton
endomorphisms contained in $\hat{\rho}$ have the same dimension and appear with the same
multiplicity.  
\eprop

Now we make the connection between Thm.\ \ref{graded2} and the case of QFT at hand.

\bdefin Let $\2F$ be a CQFT with compact global symmetry group $G$. The category
$G-\mbox{Sol}(\2F)$ is the category whose objects are transportable $G$-soliton
endomorphisms with finite index and all finite direct sums of them (not necessarily
corresponding to the same $g\in G$). The morphisms are the intertwiners in $\2F$. 
\edefin

\bprop \cite{mue11}
The category $G-\mbox{Sol}(\2F)$ is a crossed $G$-catgory in the sense of Defin.\
\ref{graded1}. (The action of $G$ is by $\rho\mapsto \alpha_g\circ\rho\circ\alpha_g^{-1}$
on the objects and by $s\mapsto\alpha_g(s)$ on the morphisms $s\in\2F$.) 
\eprop

\btheor \cite{mue11} \label{FtoA}
Consider a pair $(\2F,G; \2A,\2S)$ with $G$ finite. Then we have the equivalences
\bean G-\mbox{Sol}(\2F) & \simeq & Rep(\2A)\rtimes\2S, \\
   Rep(\2A) & \simeq & [G-\mbox{Sol}(\2F)]^G \eean
of braided $G$-crossed tensor categories and braided tensor categories, respectively.
\etheor

The situation can be neatly summarized in the following diagram, where the horizontal
inclusions of categories are full. (A very similar diagram appeared in \cite{mue1} in a
massive context where, however, one has to do with partially broken quantum symmetries.)
\begin{diagram}
 \null & \null & \lTo^{0-\mbox{grade}} & \null & \null \\
 \null & Rep(\2F) & \subset & $G$-\mbox{Sol}(\2F) & \null \\
 \uTo^{\cdot\rtimes\2S} & \cup & & \cup & \dTo_{G-\mbox{fixpoints}} \\
  \null & Rep(\2A)\cap\2S' & \subset & Rep(\2A) 
\end{diagram}\\

In view of these results it is clearly desirable to know which soliton endomorphisms a
theory $\2F$ with global symmetry $G$ admits. This is partially answered by the following
result. We say that $\2F$ admits $g$-soliton endomorphisms if for every $\2O\in\2K$ there
is an irreducible $g$-soliton endomorphism associated with $\2O$.

\bcoro \label{solis2}
Let $\2F$ be completely rational and let $\alpha_g$ be a global symmetry of finite
order, i.e. $\alpha_g^N=\id$ for some $N\in\7N$. Then $\2F$ admits $g$-soliton
endomorphisms. \ecoro  
\prf Let $G$ be the finite cyclic group generated by $\alpha_g$ and $\2A=\2F^G$. 
By Thm.\ \ref{A_comp_ratl} and Thm.\ \ref{KLM_main} we have $\2Z(Rep(\2A))=\11$. 
Then Thm.\ \ref{graded2} implies that the grading of 
$G-\mbox{Sol}(\2F)\simeq Rep(\2A)\rtimes\2S$ is full.
\qed

\brem 1. Now we can complete the proof of the implication $\2A$ completely rational 
$\impl\ \2F$ satisfies split (thus complete rationality). By Thm.\ \ref{KLM_main},
$Rep(\2A)$ is modular, thus the grading of $Rep(\2A)\rtimes\2S$ is full by Thm.\
\ref{graded2}. Let $I,J\in\2I$ satisfy $\ol{I}\cap\ol{J}=\emptyset$ and
$x\not\in\ol{I\cup J}$. By Thm.\ \ref{FtoA} (whose proof does not assume complete
rationality of $\2F$!) $\2F\restr S^1-\{x\}$ admits $g$-soliton endomorphisms for all
$g\in G$. Using the latter one can construct a normal conditional expectation 
\[ m_2: \2F(I)\vee\2F(J) \longrightarrow \2F(I)\vee\2A(J). \]
The rest of the proof works as in \cite[Sect.\ 5]{dopl}. \\
2. If $\mu_2(\2F)=1$ one can prove that $\2F$ admits soliton {\it auto}morphisms, see
the next subsection. We expect that there are direct proofs of Coro.\ \ref{solis2} and the
above fact which avoid the detour through the fixpoint theory and its modularity and 
thus might work without the periodicity restriction. 
\erem

Putting everything together we have the following generalization of Thm.\ \ref{KLM_main}:
\btheor Let $\2F$ be a completely rational CQFT with finite group $G$ of global
symmetries. Then $G-Sol(\2F)$ is a modular crossed $G$-category in the sense of \cite{t2}
with full grading.
\etheor

We end this section with the computation of $Rep(\2A)$ in a relatively simple albeit 
non-trivial and instructive example.

\subsection{An Example: Holomorphic Orbifold Models}\label{orbif}
In order to illustrate the computation of the representation category of a fixpoint
theory we consider the simplest possible example, namely the case where the net $\2F$
is completely rational with $\mu_2(\2F)=1$, i.e.\ without non-trivial sectors. Even though 
the analysis of this particular case reduces essentially to an exercise in low dimensional
group cohomology it is quite instructive and allows us to clarify, prove and extend the
results of the heuristic discussion in \cite{dvvv} and to expose the link with \cite{dpr}
and -- to a lesser extent -- with \cite{dw}. 

By the analysis in Subsect.\ \ref{rep-A} we know that the grading of $G-Sol(\2F)$ is full, 
thus $\dim [G-Sol(\2F)]_g\ge 1\ \forall g\in G$. Together with 
$\dim G-Sol(\2F)=|G|\cdot\dim Rep(\2F)=|G|$ this clearly implies that each of the
categories $[G-Sol(\2F)]_g$ has exactly one isomorphism class of simple objects, all of
dimension one and thus invertible. Therefore $G-Sol(\2F)$ is (equivalent to) the monoidal
category $\2C(G,\Phi)$ determined (up to equivalence) by $G$ and $\Phi\in H^3(G,\7T)$,
which is considered in \cite[Chap.\ 7.5]{fk} and \cite[Ex.\ 1.3]{t2}. (In our field
theoretic language this means that 
$\2F$ admits $G$-soliton {\it auto}morphisms which are unique up to inner unitary
equivalence. In an operator algebraic setting it is long known \cite{suth} that
`$G$-kernels', i.e.\ homomorphisms $G\rarr\mbox{Out}\2M=\mbox{Aut}\2M/\mbox{Inn}\2M$ are
classified by $H^3(G,\7T)$ if $\2M$ is a factor. This analysis is immediately applicable
to the present approach to CQFTs.) Now, in \cite{dpr} starting from a finite group $G$ and
a 3-cocycle $\phi\in Z^3(G,\7T)$ a quasi-Hopf algebra $D^\phi(G)$ was defined, the
`twisted quantum double'. For the trivial cocycle this is just the ordinary quantum
double. For cocycles in the same cohomology class the corresponding twisted quantum
doubles are related by a twist of the coproduct which induces an equivalence as rigid
braided tensor categories of the representation categories.

To make a long story short we state the following result:
\btheor \cite{mue11} Let $\2F$ be completely rational with $\mu_2(\2F)=1$, let $G$ be a
finite group of symmetries and let $\2A=\2F^G$. Then there is $\Phi\in H^3(G,\7T)$ such
that the following equivalences of braided crossed $G$-categories and braided categories,
respectively, hold:
\bean G-Sol(\2F) &\simeq &\2C(G,\Phi), \\
   Rep(\2A) &\simeq & D^\phi(G)-mod, \eean
where $[\phi]=\Phi$. In particular, $Rep(\2A)$ and $D^\phi(G)-\mod$ give rise to the same
representation of $SL(2,\7Z)$. 
\etheor

The proof proceeds by explicitly constructing (typically reducible if $G$ is non-abelian)
endomorphisms of $\2F$ as direct sums of soliton automorphisms and considering their
restriction to $\2A$. One finds enough inequivalent irreducible sectors of $\2A$ to
saturate the bound $\dim Rep(\2A)\le |G|^2$ and concludes that there are no others. Then
modularity of $Rep(\2A)$ follows by an easy argument based on \cite[Coro.\ 4.3]{mue4}
even without invoking the main theorem of \cite{klm}.

We conclude this discussion by emphasizing that the above analysis, satisfactory as it is,
holds only if $\2F$ is a local, i.e.\ purely bosonic, theory. If $\2F$ is fermionic
(graded local) then new phenomena may appear, as is illustrated by the following well
known example \cite{boc2}. Let $\2F$ be a theory of $N$ free real fermions on $S^1$ and
let $\2A=\2F^G$ where $G=\7Z/2$ acts by $\psi\mapsto -\psi$. $\2F$ satisfies twisted
duality for all disconnected intervals $E$, thus $\mu_2(\2F)=1$ (in a generalized sense)
and $\mu_2(\2A)=4$. One finds $|H^3(G,\7T)|=2$, corresponding to the fusion rules 
$\7Z/2\times \7Z/2$ and $\7Z/4$ for $D^\phi(G)$, which in fact govern the cases $N=4M$ and
$N=4M-2$, respectively. For odd $N$, however, $Rep(\2A)$ has only three simple objects
(of dimensions $1, 1, \sqrt{2}$) with Ising fusion rules \cite{boc1}. The latter case
clearly is not covered by \cite{dvvv} and our elaboration \cite{mue11} of it. While the
appearance of the object with non-integer dimension can be traced back to the fact that
$\2F$ does not admit soliton automorphisms for $g\ne e$ but rather soliton endomorphisms,
a general model independent analysis of the fermionic case is still lacking and seems
desirable.

\section{Summary and Open Problems}
At least on an abstract level the relation between the representation categories of
rational CQFTs $\2F$ and $\2A=\2F^G$ for finite $G$ has been elucidated in quite a
satisfactory way by Thm.\ \ref {AtoF} and Thm.\ \ref{FtoA}. We have seen that this leads
to fairly interesting structures, results and conjectures of an essentially categorical
nature. When considering concrete QFT models the computations can, of course, still be
quite tedious as is amply demonstrated by \cite{x2} and \cite{mue11}. \\ 

We close with a list of important open problems.
\begin{enumerate}
\item Extend the results from Props.\ \ref{sce1}, \ref{sce2} to extensions $\2C\rtimes\2S$
where $G$ is non-abelian. Thus, (i) given a simple object $X\in\2C$, understand how
$F(X)\in\2C\rtimes\2S$ decomposes into simple objects. (ii) Clarify the structure of the
set of isomorphism classes of simple objects in $\2C\rtimes\2S$. (iii) Compute the fusion
rules of $\2C\rtimes\2S$ and the $S$-matrix of $(\2C\cap\2S')\rtimes\2S$.
\item Prove a form of unique factorization for modular categories into prime ones.
\item Prove Conjecture \ref{minext} on the existence of minimal modular extensions.
\item Give a more direct proof of Coro.\ \ref{solis2} on the existence of soliton
endomorphisms. 
\end{enumerate}

We cannot help remarking that the results of our Categorical Interludes strongly resemble
well-known facts in Galois theory and algebraic number theory. (Note, e.g., the striking
similarity between our Prop.\ \ref{crucial} and Coro. 2-3 of \cite[\S I.7]{lang} on the
decomposition of prime ideals in Galois extensions of quotient fields of Dedekind rings,
thus in particular algebraic number fields.) The same remark applies to questions 1-3
above.



\begin{thebibliography}{99}
\bibitem{mue11} M. M\"uger: Global symmetries in conformal field theory: Orbifold
   theories, simple current extensions and all that. In preparation.
\end{thebibliography}
\end{document}